# 1T'-MoTe$_2$ single crystal: a possible topological superconductor

X. Luo[1]*, F. C. Chen[1,2], J. L. Zhang[3], Q. L. Pei[1], G. T. Lin[1,2], W. J. Lu[1], Y. Y. Han[3], C. Y. Xi[3], W. H. Song[1] and Y. P. Sun[3,1,4*]

[1] Key Laboratory of Materials Physics, Institute of Solid State Physics, Chinese Academy of Sciences, Hefei, 230031, China

[2] University of Science and Technology of China, Hefei, 230026, China

[3] High Magnetic Field Laboratory, Chinese Academy of Sciences, Hefei, 230031, China

[4] Collaborative Innovation Center of Advanced Microstructures, Nanjing University, Nanjing, 210093, China

## Abstract

We measured the magnetoresistvity (MR) properties of the 1T'-MoTe$_2$ single crystal under the magnetic field up to 33 T. By analyzing the Shubnikov–de Haas oscillations of MR at the low temperature, single Fermi surface is revealed. From the strong oscillatory component of the longitudinal resistance $\Delta R_{xx}$, a linear dependence of the Landau index $n$ on $1/B$ is obtained. The intercept of the Landau index plot is between 3/8 and 1/2. This clearly reveals a nontrivial $\pi$ Berry's phase, which is a distinguished feature of Dirac fermions. Accompanied by the superconductivity observed at $T_C$=0.1 K, 1T'-MoTe$_2$ may be a promising candidate of the topological superconductor.

Corresponding author: xluo@issp.ac.cn and ypsun@issp.ac.cn.

The theoretical prediction and successful experimental on the topological insulators have opened an exciting research door in the physical and materials fields.[1-5] Those materials have attracted tremendous interest not only in the potential applications but also in searching for the new topological phase. For example, the topological superconductors are new families of materials with novel electronic states and have a full pairing gap in the bulk and a topological protected gapless surface state.[6] Due to the unique electron structure, the topological superconductors may be used in topological quantum computing in future.

Two dimensional (2D) transition-metal dichalcogenides (TMDs) materials, named $MX_2$ where M is a transition metal (Ta, Nb, Mo and W) and X is a chalcogen (S, Se and Te). $MX_2$ materials have attracted renewed interest owing to their rich physical properties, such as quantum spin Hall (QSH) effect, Weyl Semimetal (WSM), superconductivity, unsaturated magnetoresistivity (XMR) and so on.[7-11] Recently, $MoTe_2$ has attracted much attention because it was predicted as a new type-II WSM candidate.[11] $MoTe_2$ crystallizes into three different phase: $2H$, $1T'$ and $T_d$ phases. The $2H$ phase presents a semiconducting behavior. However, the $1T'$ and $T_d$ phases show semi-metallic and exhibit pseudo-hexagonal layers with zig-zag metal chains. The $T_d$ compound can be obtained by cooling the $1T'$ phase down to 240 K.[12, 13] Very recently, the $T_d$-$MoTe_2$ is reported to be a superconductor with $T_C$=0.1 K and the small pressure applied or the partial Te ion substitution by S one can dramatically enhance the $T_C$ and a dome-shaped superconducting phase diagram is observed.[9, 14] 1T'- $MoTe_2$ is believed to be a promising candidate of topological superconductor in the bulk materials.[9] Although numerous electronic structure calculations of $MoTe_2$ have been conducted, the observation of quantum oscillations, which are widely studied to resolve the electronic structure of topological materials, is still missing in $MoTe_2$. For a topological superconductor, it is necessary to observe quantum oscillations to resolve Landau level quantization. And another distinguished feature of Dirac fermions is the nontrivial $\pi$ Berry's phase which results from their cyclotron motion.[15–17] The nontrivial $\pi$ Berry's phase is a geometrical phase factor and it is acquired when an electron circles a Dirac point and can be experimentally accessed by analyzing the Shubnikov-de Haas ($SdH$) oscillations. The $SdH$ has already been employed in studies of $SrMnBi_2$, $Sr_xBi_2Se_3$, $Cd_3As_2$, LaSb, topological insulators and so on.[18-23]

Herein, we present the quantum oscillation study of the MR for 1T'-MoTe2 single crystal under the high magnetic field up to 33 T. By analyzing the *SdH* oscillations of MR at low temperature, single Fermi surface is revealed. At the same time, a nontrivial $\pi$ Berry's phase, which is a distinguished feature of Dirac fermions, is observed in 1T'-$MoTe_2$ single crystal. Meanwhile, the superconductivity with $T_C$=0.1 K in 1T'-$MoTe_2$ has been reported. Therefore, $MoTe_2$ single crystal may be a promising candidate of the topological superconductor.

1T'-$MoTe_2$ single crystals were grown by the flux method. Mo (Alfa Aesar, 99.9 %) and Te (Alfa Aesar, 99.9 %) powders with a stoichiometric ratio were mixed with sodium chloride (NaCl, Alfa Aesar, 99.9 %) in an alumina crucible, which was sealed in a quartz tube under vacuum at a high pressure of $10^{-5}$ Torr. The crystal growth recipe followed the reported paper.[7] Flake-like crystals with shinning surfaces were obtained. The size of the crystal was about 2*1*0.1 $mm^3$. The single crystals were air-stable and can be easily exfoliated. Powder X-ray diffraction (XRD) patterns were taken with Cu $K_{a1}$ radiation ($\lambda$=0.15406 nm) using a PANalytical X'pert diffractometer at room temperature. The element analysis of the single crystals was performed using a commercial energy dispersive spectroscopy (EDS) microprobe. The element compositions of the single crystals used in the text are the nominal ones. The electrical transport measurements were performed by Physical Properties Measurement System (PPMS-16 T). High-field MR oscillation was performed at the High Magnetic Field Laboratory of the Chinese Academic Sciences (water cooling magnet WM-5, H=33 T).

The $MoTe_2$ crystallizes into three different phases: *2H*, *1T* and $T_d$ phases, as shown in Fig. 1 (a). The $T_d$ phase can be obtained by cooling the 1T' phase down to 240 K.[12, 13] Figure 1 (b) shows the powder X-ray diffraction (XRD) patterns of the crushed 1T'-$MoTe_2$ single crystals. Detailed refinement of XRD patterns suggest that the lattice parameters of 1T'-$MoTe_2$ are $a$=3.4715 Å, $b$=6.3382 Å, $c$=13.8611 Å and $\beta$=93.75°, which are consistent with the reported results.[9, 12, 13] The 1T'-$MoTe_2$ single crystal shows a flake-like shape, the largest natural surface is *ab* plane and the longest axis is *a* direction. As shown in the inset of Fig. 1 (d), the quality of the 1T'-$MoTe_2$ single crystals was further checked by the X-ray rocking curve and the full width at half maximum (FWHM) of the (004) Bragg peak is only 0.07°, indicating the high quality of the single crystals. The crystal was cleaved, with 2*1 $mm^2$ in the *ab* plane and a 0.1 mm thickness. A standard six-probe method was used for both the longitudinal resistivity and transverse Hall resistance measurements. A magnetic field was applied perpendicular to the *ab* plane up to 33 T and the current was along the *a* axis.

Figure 2 (a) shows the temperature dependence of the longitudinal resistivity $\rho_{xx}$ at H=0 T, 8.5 T and 16 T for a 1T'-MoTe$_2$ single crystal. The current direction is along the Molybdenum chains (*a* axis) and the magnetic field is applied perpendicular to the dichalcogenide layers, along the *c* axis. The temperature dependent resistivity under various applied magnetic fields (H up to 16 T) is shown in Fig. 2 (a). Below 10 K, the $\rho$(T) curve at H=0 T is very flat, extrapolating to a residual resistivity of $\rho_{xx0}$=11.3 μΩ cm. The electrical resistivity data show a metallic behavior (*dρ/dT>0*) with the large residual resistivity ratio (RRR) of $\rho_{300 K}/\rho_{2 K}$~256, which indicates that the 1T'-MoTe$_2$ single crystal is of high quality. As shown in Fig. 2 (a), an anomaly with a hysteresis in the *ρ(T)* of MoTe$_2$ single crystals is observed around 240 K, which corresponds to the structural phase transition.[12, 13] When a field is applied, the resistivity of the single crystal essentially follows the zero-field curve until it is cooled close to the 'turn on' temperature $T_{MI}$ (identified as the minimum in the resistivity). Once below $T_{MI}$, the resistivity begins to increase markedly, which is similar to the behavior of WTe$_2$.[10] The $T_{MI}$ temperature shifted to a higher temperature under the higher magnetic field with a rate of 1.33 K/T. Figure 2 (b) presents the Hall resistivity $\rho_{xy}$ as a function of magnetic field at various temperatures. The negative slope of $\rho_{xy}$ means the dominant charge carriers in 1T'-MoTe$_2$ are electrons, and the carrier concentration of $n_e$≈2.3*10$^{20}$ cm$^{-3}$ and a mobility of 3900 cm$^2$/V s are estimated from the low-field slope, which are consistent with the reported results.[7] The low concentration and high mobility of charge carriers in 1T'-MoTe$_2$ single crystal also indicate the high quality of present studied crystals. In Fig. 2 (c), there are clear oscillations of $\rho_{xx}$. The MR effect at low temperature of T=2 K is very huge and without saturation. The MR is defined by MR $= \frac{[\rho_{33 T} - \rho_{0 T}]}{\rho_{0 T}} \times 100$ % ≈61700 % in a field of 33 T. This MR is quadratic for low field and almost linear for larger *B* without saturation, similar to many known semimetallic materials such as WTe$_2$, TaAs(P), NbAs(P) and so on. [10, 23-26]

As shown in Fig. 3 (a), the MR increases as temperature decreases. At 2 K, the MR oscillations can be clearly tracked as H≈9 T. The MR oscillations are not observed when the temperature is above 6 K. After subtracting a smooth background, Fig. 3 (b) shows the oscillatory component of *ΔR$_{xx}$* versus *1/B* at various temperatures. *ΔR$_{xx}$* are periodic in *1/B*, as expected from the successive emptying of the Landau levels when the magnetic field is increased. Figure 3 (c) shows a single oscillation frequency of $F$=242 T obtained from the fast Fourier transform (FFT) spectra at $T$=2 K, which corresponds to $\Delta(1/B)$=0.00413 T$^{-1}$.

According to the Onsager relation:

$$F = \frac{\hbar}{2\pi e} A(\epsilon_F),$$  (1)

The frequency of the *SdH* oscillations $F$ is proportional to $A(\varepsilon_F)$ which is the cross section of the Fermi surface in the plane perpendicular to the applied magnetic field. $\hbar$ is Planck's constant and $e$ is the electron charge. Substituting the average frequency $F$=242 T in Eq. (1) and assuming a circular Fermi surface, we obtain the Fermi momentum $k_F$=0.0858 Å.

The *SdH* oscillation amplitude can be further described by the Liftshitz-Kosevich formula: [27, 28]

$$\Delta R(T, B) \propto \frac{\frac{2\pi^2 k_B T}{\Delta E_N(B)}}{\sinh\left(\frac{2\pi^2 k_B T}{\Delta E_N(B)}\right)} e^{-\frac{2\pi^2 k_B T_D}{\Delta E_N(B)}} cos2\pi \times (\frac{F}{B} + \frac{1}{2} + \beta)$$  (2)

Where $\Delta E_N(B)$=$\hbar eB/m^*$is the energy gap between the *Nth* and *(N+1)th* Landau levels, $T$ and $B$ are temperature and magnetic field, $T_D$ is the Dingle temperature, $m^*$ is the effective mass of the carriers, and $\hbar$ and $k_B$ are the Planck's constant and the Boltzmann constant, respectively. $2\pi\beta$ is the Berry's phase and will be discussed later. In order to further investigate on the physical properties of 1T'-MoTe$_2$, we did analysis on the $\Delta R_{xx}$ data shown in Fig. 3 (b). The temperature dependence on the normalized oscillation amplitude at $1/B$=0.06873 T$^{-1}$, which corresponds to the eighteenth Landau level, is shown in Fig. 3 (d). The corresponding results indicates that the effective mass of the carriers is 0.72 $m_0$ and Fermi velocity is $v_F$=1.32*10$^5$ m/s (where $m_0$ is the effective mass of a free electron). The means-free path along the *a* axis is about 102 nm using $\mu = e\ell/\hbar k_F$, where $\mu$ is the carrier mobility, $\ell$ is the mean-free path, $\hbar$ and $k_F$ are the Planck's constant and the Fermi momentum, respectively. The mean-free path $\ell$ is two orders of magnitude shorter than that of WTe$_2$.[29] From the formula $\ell = v_F \tau_S$, where $v_F$ is the Fermi velocity and $\tau_S$ is the scattering time. The obtained $\tau_S$ is about 7.3*10$^{-13}$ s. The obtained parameters are summarized in **Table I**.

Concerning the properties of the Berry's phase in 1T'-MoTe$_2$, Fig. 4 (a) shows the magnetic field dependence on $\Delta R_{xx}$, which is obtained from the *R(T)* (shown in Fig. 2 (c)). $\Delta R_{xx}$ is achieved through subtracting the background of the *R(T)* data under the high magnetic field up to 33 T. Figure 4 (b) shows the Landau index plot, *n* versus *I/B*. We assign integer indices to the $\Delta R_{xx}$ valley positions in $1/B$ and half integer indices to the $\Delta R_{xx}$ peak positions.[18, 23] According to the Lifshitz-Onsager quantization rule $A_F \left( \frac{\hbar}{eB} \right) = 2\pi(n + \frac{1}{2} + \beta + \delta)$, the Landau index $n$ is linearly dependent on $1/B$. $2\pi\beta$ is the Berry's phase, and the $2\pi\delta$ is an additional phase shift as a result of the curvature of the Fermi surface in the third direction.[28] The value of $\delta$ changes from 0 for a quasi-2D cylindrical Fermi surface to $\pm 1/8$ for a corrugated 3D Fermi surface. As shown in Fig. 4 (b), all the data fall into a straight line and the liner extrapolation has an intercept 0.47. As we know, for a conventional metal with the trivial parabolic dispersion, the Berry's phase $2\pi\beta$ should be zero. However, for a Dirac system with linear dispersion, there should be a nontrivial $\pi$ Berry's phase ($\beta=1/2$). The $\pi$ Berry's phases have been observed in 2D graphene, $SrMnBi_2$ and $Sr_xBi_2Se_3$, BiTeI and so on.[18-23] The $\pi$ Berry's phase is clearly revealed by the intercept 0.47 shown in Fig. 4 (b), and thus provide strong evidence for the existence of Dirac fermions in 1T'-$MoTe_2$ single crystal. The slight deviation from $\beta=1/2$ indicates an additional phase shift $\delta \approx 0.03$, which may result from the quasi-2D nature of the system. We also measured the angle dependence of the MR under the magnetic field up to 33 T and conducted the FFT, as shown in Fig. 1S. We can successfully obtain the reliable data at the low degree zone. The angle dependence of the FFT frequency does roughly follow the $1/cos(\theta)$ (the red dash-dot line shown in Fig. 2S), which is usually used to describe the 2D Fermi surface. It means the quasi-2D nature of the Fermi surface may exist in the 1T'-$MoTe_2$ single crystal. With its superconductivity at $T_C$=0.1 K is observed,[9] the 1T'-$MoTe_2$ is a possible topological superconductor candidate. However, further experiments, such as angle-resolved photoemission spectroscopy (ARPES) and scanning tunneling microscopy (STM) at low temperatures, are needed to investigate the structure of the Fermi surface in 1T'-$MoTe_2$ single crystals in future.

Based on the calculations of the electronic structure performed on $MoTe_2$, it seems that the band structure of $MoTe_2$ is more complex than the simple linear energy dispersion near the Fermi surface obtained from the present quantum MR measurements.[11] Now, we try to analyze the possible reasons. Firstly, we focus on the well-studied 1T'-$MoTe_2$ single crystals. As we know, through different sintering recipes, the phase transition between the 2H- and 1T'-$MoTe_2$ single crystals can be achieved, where the Te vacancies exert an important effect on the phase transition.[7, 8] *Cho et al.* also reported the Te defects or vacancies affect on the transport properties of MoTe$_2$ single crystals.[8] For the present MR measurements, we did the cleaving before the measurements, because the van der Waals force exists between the Te-Mo-Te layers and the cleaving surface is usually the Te-Mo-Te layer. So the Te vacancies may also affect the quantum oscillation measurements. We did the EDS to analyze the real composition of the MoTe$_2$ single crystals. The results are shown in the **Table I** of the supporting materials. We did the measurements on the five different positions on the present studied crystal. It shows that the practical composition of Te element does not match the nominal one and nearly 2.5~5 % of Te one are lost for the crystal. As we know, the Te vacancies induced by the laser irradiation can also dramatically change the electrical properties in MoTe$_2$ single crystals.[8] Therefore, the difference between the Dirac-like Fermi surface obtained from the present quantum oscillations study and the calculated one may be related to the loss of Te element in 1T'-MoTe$_2$ single crystals. Secondly, the reported calculations have been done without the magnetic field. However, the strong spin-orbital coupling (SOC) effect can obviously change the electronic structure of MoTe$_2$.[8, 11] On the other hand, the SOC effect may be more active under the high magnetic field in MoTe$_2$. Therefore, further calculations based on the defect 1T'-MoTe$_{2-\delta}$ are needed, especially the ones under the high magnetic field are desired.

In summary, we have performed bulk quantum transport measurements on single crystals of the WSM candidate 1T'-MoTe$_2$. By analyzing the *SdH* oscillations of longitudinal resistance at low temperature, a nontrivial $\pi$ Berry's phase with a small phase shift is obtained, which provides the bulk quantum transport evidence for the existence of a Dirac semimetal phase in 1T'-MoTe$_2$. Accompanied by its superconductivity with $T_C$=0.1 K, the 1T'-MoTe$_2$ is a possible topological superconductor candidate and the unique quantum transport properties in 1T'-MoTe$_2$ may open up new avenues for future device applications.

# Acknowledgements

This work was supported by the Joint Funds of the National Natural Science Foundation of China and the Chinese Academy of Sciences' Large-Scale Scientific Facility under contracts (U1432139, U1232139), the National Nature Science Foundation of China under contracts (51171177, 11404342), the National Key Basic Research under contract 2011CBA00111, and the Nature Science Foundation of Anhui Province under contract 1508085ME103, 1408085MA11. The authors thank Dr. Chen Sun for her assistance in editing the manuscript.

**Table I:** The obtained physical parameters in 1T'-MoTe$_2$ single crystal

| | 1T'-MoTe$_2$ |
|---|---|
| $F_S$ | 242 $T$ |
| $A$ | 0.2312 $nm^{-2}$ |
| $n_{exp.}$ | 2.3*10$^{20}$ $cm^{-3}$ |
| $n_{Hall}$ | 3900 $cm^2/V\,s$ |
| $k_F^x$ | 0.858 $nm^{-1}$ |
| $m$ | 0.72 $m_0$ |
| $v_F$ | 1.32*10$^5$ $m/s$ |
| $\tau_F$ | 7.3*10$^{-13}$ $s$ |
| $l$ | 105 $nm$ |

**Figure 1:**

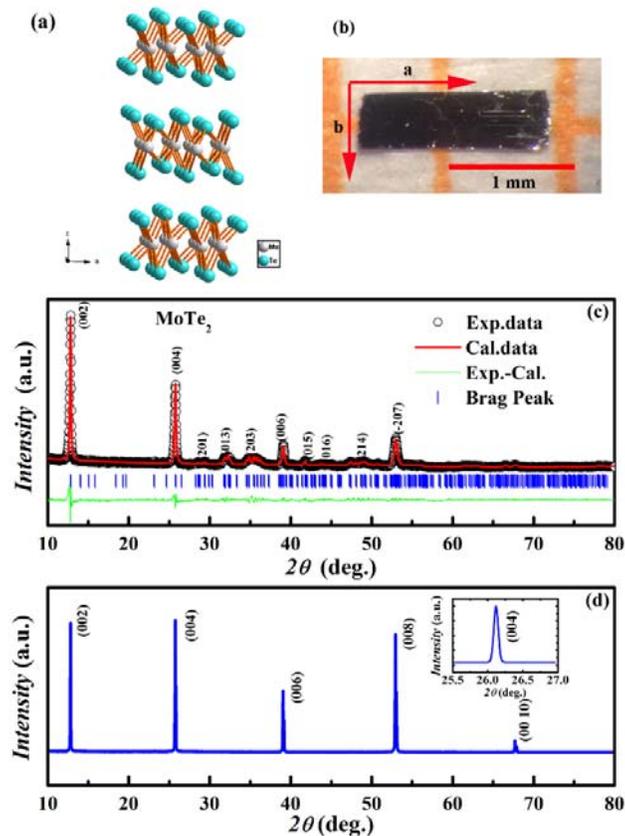

**Fig. 1 (color online): (a)** The crystal structure of $T_d$-MoTe$_2$; **(b)** The picture of studied MoTe$_2$ single crystal. The crystal size is approximately 2*0.5*0.1 mm; **(c)** Rietveld refined powder XRD patterns at room temperature for the crushed MoTe$_2$ crystals. The vertical marks (blue bars) stand for the position of Bragg peaks, and solid line (green line) at the bottom correspond to the difference between experimental and calculated intensities; **(d)** XRD patterns of the crystal measured on the (001) surface. Inset presents a typical X-ray rocking curve of the (004) Bragg peak.

**Figure 2:**

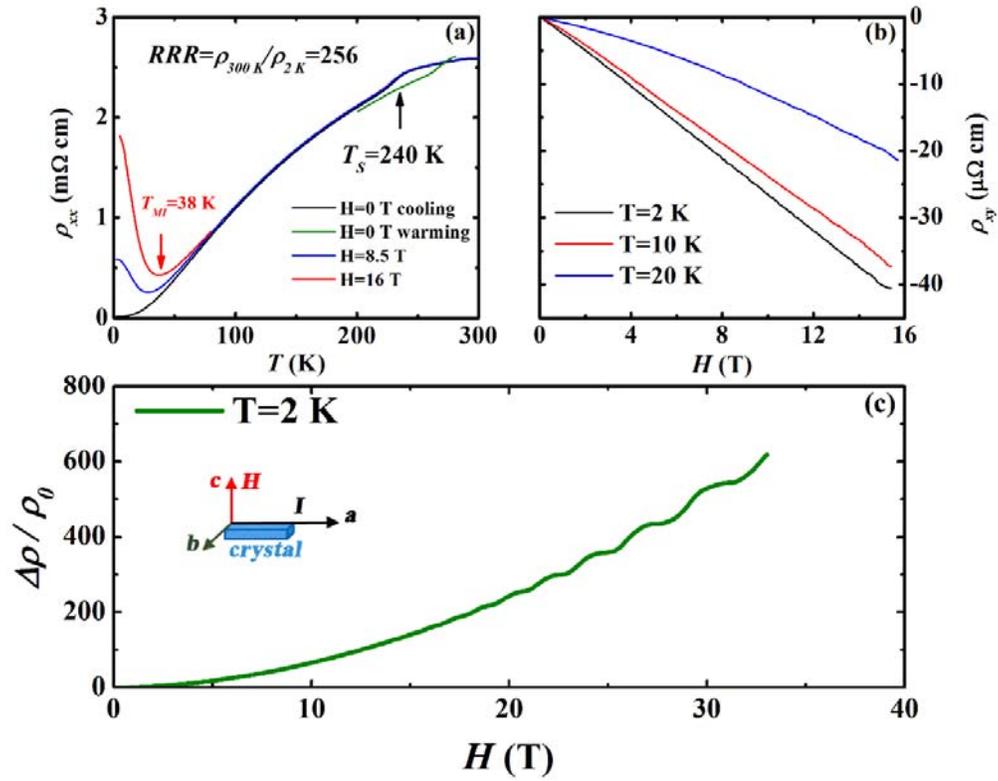

**Fig. 2 (color online):** **(a)** The temperature dependence of resistivity under different magnetic fields. $T_{MI}$ is defined as the temperature at which the resistivity is a minimum, where the MR is turned on; **(b)** The Hall resistivity $\rho_{xy}$ at 5, 10 and 50 K; **(c)** The *SdH* Oscillations of longitudinal MR at *T*=2 K with the field up to 33 T. The inset shows the configuration of the resistivity measurement (the current flows in the *ab* plane along the *a* axis.).

**Figure 3:**

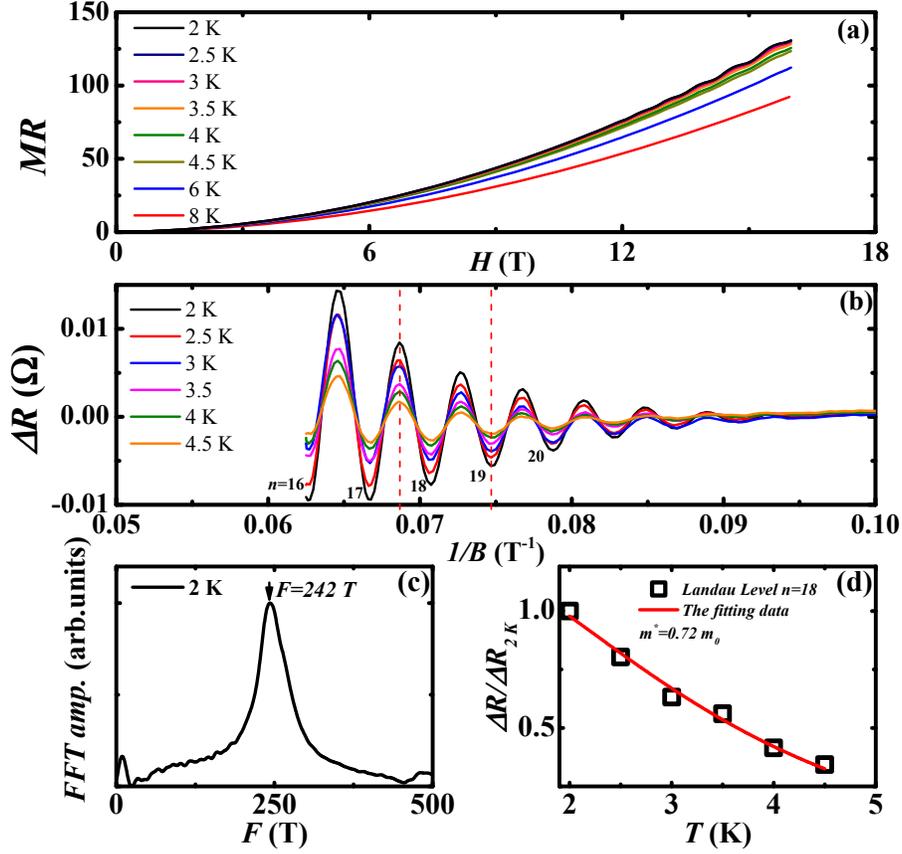

**Fig. 3 (color online): (a)** The temperature dependence of the MR at different temperatures; **(b)** The oscillatory component $\Delta R_{xx}$, extracted from $R_{xx}$ subtracting a smooth background; **(c)** A single oscillation frequency $F$=242 T identified from fast Fourier transform spectra at $T$=2 K; **(d)** The temperature dependence of the relative amplitude of $\Delta R_{xx}$ for the eighteenth Landau level. The solid line is a fit to the Lifshitz-Kosevich formula, from which we get the cyclotron effective mass $m^* \approx 0.72 \ m_0$ and Fermi velocity $v_F$=1.32*10$^5$ m/s.

**Figure 4:**

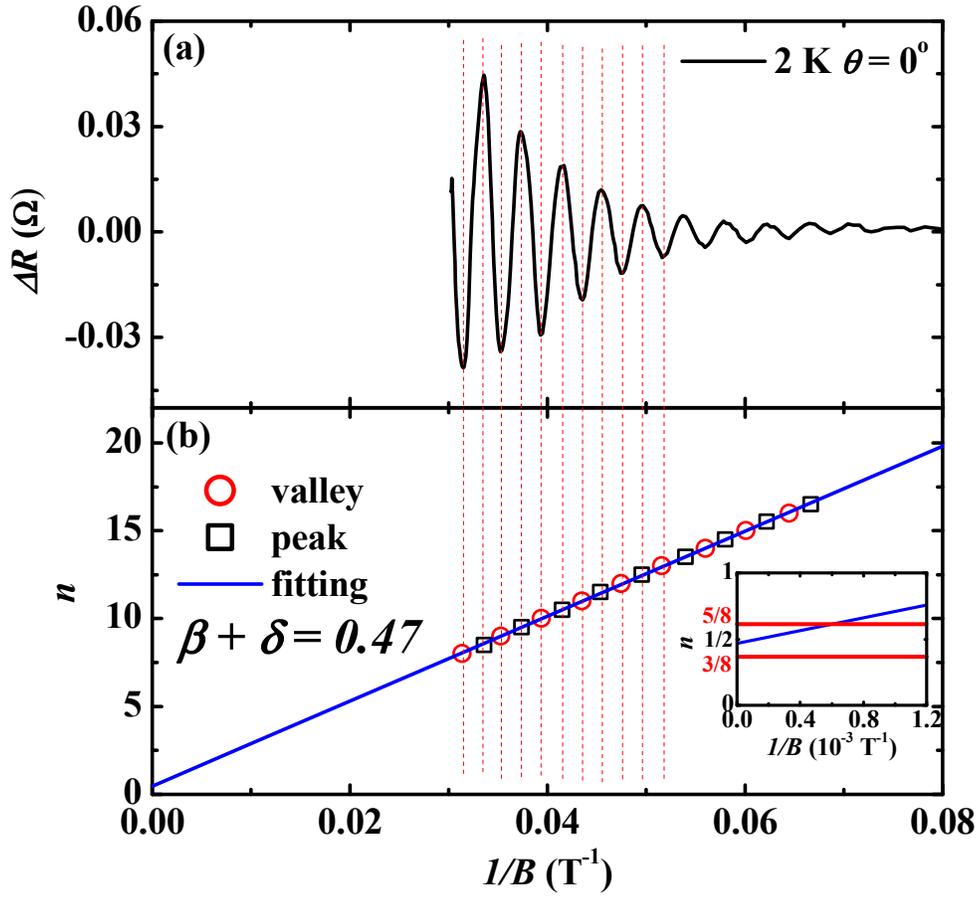

**Fig. 4 (color online):** **(a)** The oscillatory component $\Delta R_{xx}$, extracted from $R_{xx}$ (T=2 K and the magnetic field up to 33 T) by subtracting a smooth background; **(b)** Landau index $n$ plotted against $1/B$. The closed circle denote the integer index (the valley of $\Delta R_{xx}$), the open circle indicate the half integer index (the peak of $\Delta R_{xx}$). The left inset shows the intercept of $\Delta R_{xx}$ lie between $\beta$ and $\beta \pm 1/8$ ($\beta$=1/2), which is strong evidence for a non-trivial $\pi$ Berry's phase of Dirac fermions in MoTe$_2$.